\documentclass[aps,showpacs,superscriptaddress,preprint,
nofootinbib,tightenlines]{revtex4-1}

\usepackage[utf8]{inputenc}
\usepackage[english]{babel}
\usepackage{amssymb,amsthm,amsmath,amstext,amsbsy,amsopn}
\usepackage{nicefrac}
\usepackage{slashed}
\usepackage{graphicx}
\usepackage{hyperref}
\usepackage{leftidx}
\usepackage{environ}
\usepackage{mathtools}
\usepackage{xspace}
\usepackage{suffix}
\usepackage[english]{babel}
\usepackage{hyperref}
\usepackage{dcolumn}

\usepackage{longtable}
\usepackage{float}
\usepackage[english]{cleveref}

\newcommand{\apriori}{\textit{a priori}\xspace}
\WithSuffix\newcommand\apriori*{\textit{a-priori}\xspace}

\newcommand*\rvec[1]%
{\ensuremath{\overset{\smash{\raisebox{-1.5pt}{\tiny$\rightarrow$}}}{#1}}}
\newcommand*\lvec[1]%
{\ensuremath{\overset{\smash{\raisebox{-1.5pt}{\tiny$\leftarrow$}}}{#1}}}

\NewEnviron{subalign}[1][]{%
\begin{subequations}\begin{align}
  \BODY
\end{align}\label{#1}\end{subequations}
}

\NewEnviron{spliteq}{%
\begin{equation}\begin{split}
  \BODY
\end{split}\end{equation}
}

\newcommand{\simge}{\hspace*{0.2em}\raisebox{0.5ex}{$>$}
     \hspace{-0.8em}\raisebox{-0.3em}{$\sim$}\hspace*{0.2em}}

\def\vec#1{{\bf #1}}

\newcommand{\beq}{\begin{equation}}
\newcommand{\eeq}{\end{equation}}
\newcommand{\bqa}{\begin{eqnarray}}
\newcommand{\eqa}{\end{eqnarray}}

\def\mqo2{{\!\!\!}}

\newcommand{\K}[1]{\ensuremath{\left(#1\right)}}

\newcommand{\dxi}{\Delta\left(\xi\right)}

\newcommand{\maxDeviation}{\(d_{\rm max}\)}

\newcommand{\leftValue}{\ensuremath{ \Delta\K{-\pi} }}
\newcommand{\middleValue}{\ensuremath{ \Delta\K{-\pi/2} }}
\newcommand{\rightValue}{\ensuremath{ \Delta\K{-\pi/4} }}
\newcommand{\derivativeValue}{\ensuremath{ \Delta^\prime \K{-\pi/2} }}

\newcommand{\leftDeviationShort}{\ensuremath{ |d \K{-\pi}| }}
\newcommand{\middleDeviationShort}{\ensuremath{ |d \K{-\pi/2} | }}
\newcommand{\rightDeviationShort}{\ensuremath{ |d \K{-\pi/4} | }}
\newcommand{\derivativeDeviationShort}{\ensuremath{ |d^\prime \K{-\pi/2} | }}

\newcommand{\piv}{-\frac{\pi}{4}}

\newcommand{\pida}{-\frac{3\pi}{8}}
\newcommand{\pifa}{-\frac{5\pi}{8}}

\newcommand{\ods}{\ensuremath{\mathbb{O}}} 
\newcommand{\sods}{\ensuremath{\tilde{\mathbb{O}}}} 
\newcommand{\nds}{\ensuremath{\mathbb{N}}} 
\newcommand{\tds}{\ensuremath{\mathbb{T}}} 
\newcommand{\cds}{\ensuremath{\nds \cup \tds}} 

\begin{document}

\title{More on the universal equation for Efimov states}

\author{M. Gattobigio}
\affiliation{
 Universit\'e C\^ote d'Azur, CNRS, Institut  de  Physique  de  Nice,
1361 route des Lucioles, 06560 Valbonne, France }

\author{M. G\"obel}
\affiliation{Institut für Kernphysik, Technische Universität Darmstadt,
64289 Darmstadt, Germany}

\author{H.-W. Hammer}
\affiliation{Institut für Kernphysik, Technische Universität Darmstadt,
64289 Darmstadt, Germany}
\affiliation{ExtreMe Matter Institute EMMI,
GSI Helmholtzzentrum für Schwerionenforschung GmbH,
64291 Darmstadt, Germany}

\author{A. Kievsky}
\affiliation{Istituto Nazionale di Fisica Nucleare, Largo Pontecorvo 3,
  56127 Pisa, Italy}

\date{March 13, 2019}

\begin{abstract}
Efimov states are a sequence of shallow three-body
bound states that arise when the two-body scattering length is
much larger than the range of the interaction.
The binding energies of these states are described as a function of the
scattering length and one three-body parameter
by a transcendental equation involving
a universal function of one angular variable.
We provide an accurate and convenient parametrization of this function.
Moreover, we discuss the effective treatment of range corrections in the
universal equation and compare with a strictly perturbative scheme.
\end{abstract}

\maketitle

\section{Introduction}

The interactions of nonrelativistic particles
with short-range interactions at extremely low energies
are determined primarily by their $S$-wave scattering length $a$.
If $|a|$ is much larger
than the characteristic range $l$ of their interaction, low-energy
atoms exhibit universal properties that are insensitive to the
details of the interaction potential~\cite{Braaten:2004rn}. In the two-body sector,
$a$ is the only relevant dimensionful parameter
at low energies and all observables simply scale with appropriate
powers of $a$. If $a>0$, there is one shallow two-body bound state
(the dimer) with binding energy $E_2=\hbar^2/ma^2$.
In the three-body sector, the universal properties
were first deduced by Efimov \cite{Efi71}.
The most remarkable is the existence of a sequence of three-body
bound states with binding energies $E_3^{(n)}$
geometrically spaced in the interval
between $\hbar^2/(ma^2)$ and $\hbar^2/(ml^2)$. The number of these
``Efimov states'' is roughly $\ln(|a|/l)/\pi$
if $|a|$ is large enough. In the limit
$|a|\to \infty$, there is an accumulation of infinitely many
Efimov states at threshold (the ``Efimov effect'').
The knowledge of their binding energies is
essential for understanding the energy dependence of low-energy three-body
observables. For example, Efimov states can have dramatic
effects on atom-dimer scattering if $a>0$ \cite{Efi71,BHK99}
and on three-body recombination if $a<0$ \cite{EGB99,BBH00}.

A large two-body scattering length can be obtained by fine-tuning a
parameter in the interatomic potential to bring a
real or virtual two-body state close to threshold. The fine tuning can
be provided accidentally by nature. An example is the $^4$He atom,
whose scattering length $a= 104$ \AA\ \cite{Gri00}
is much larger than the effective range $l\approx 7$ \AA.
There are two $^4$He trimer states, the ground state state \cite{Sch94}
where range corrections are significant, and the
recently observed excited state
which is almost an ideal Efimov state \cite{Kun15}.
For atoms the fine tuning can also be obtained by
using Feshbach resonances~\cite{Chi10}.
The scattering length of alkali atoms can be changed experimentally
by tuning an external magnetic field. An important difference from He is that the
interatomic potentials of alkali atoms support many deep two-body bound states.
Using Feshbach resonances, Efimov states have been observed
in a variety of atoms including $^{133}$Cs, $^6$Li, $^7$Li, and several
mixtures of atoms (see Refs.~\cite{Bra07,Naidon:2016dpf} for reviews).

Efimov's universal equation for the energies $E_3^{(n)}$
follows from the approximate scale-invariance at
length scales $R$ in the region $l \ll R \ll |a|$ and the
conservation of probability.
Introducing polar variables $H$ and $\xi$ in the plane
defined by the variables $1/a$ and $K={\rm sgn}(E)|mE|^{1/2}/\hbar$,
the binding energies of the Efimov states are solutions to
a transcendental equation involving a single universal function
$\Delta(\xi)$ of $\xi$ \cite{Efi71}. In Ref.~\cite{Braaten:2002sr},
this equation was extended to the case where deeply bound dimers
are present by introducing a loss parameter $\eta_*$.

Here, we restrict ourselves to the case without deep dimers
and  consider the
equation for the radial wave function $f(R)$ in the adiabatic
hyperspherical representation of the three-body problem \cite{Jen93,Nie01}.
The hyperspherical radius for three identical bosons with coordinates
$\vec{r}_1$, $\vec{r}_2$, and $\vec{r}_3$ is $R^2=(r_{12}^2+r_{13}^2+
r_{23}^2)/3$, where $r_{ij}=|\vec{r}_i -\vec{r}_j|$.
If $|a| \gg l$, the radial equation for 3 bosons
with total angular momentum zero reduces in the region $l \ll R
\ll |a|$ to
\beq
-{\hbar^2 \over 2m} \left[ {\partial^2 \over \partial R^2}
+ {s_0^2 + 1/4 \over R^2} \right] f(R) = E f (R),
\label{radial}
\eeq
where $s_0 \approx 1.00624$.
This has the same form as the one-dimensional Schr{\"o}dinger
equation for a particle in an attractive $1/R^2$ potential.
If we impose a boundary condition on $f(R)$
at short-distances of order $l$, the radial equation
(\ref{radial}) has solutions at discrete negative values of the
eigenvalue $E=-E_3^{(n)}$, with $E_3^{(n)}$
ranging from order $\hbar^2/(m l^2)$ to order
$\hbar^2/(ma^2)$. The corresponding eigenstates are called Efimov states.
As $|a|\to \infty$, their spectrum approaches the simple
law $E_3^{(n)}/ E_3^{(n+1)}= e^{2\pi/s_0}\approx 515\,$.

\begin{figure}[H]
  \centering
    \includegraphics[width=0.5\textwidth]{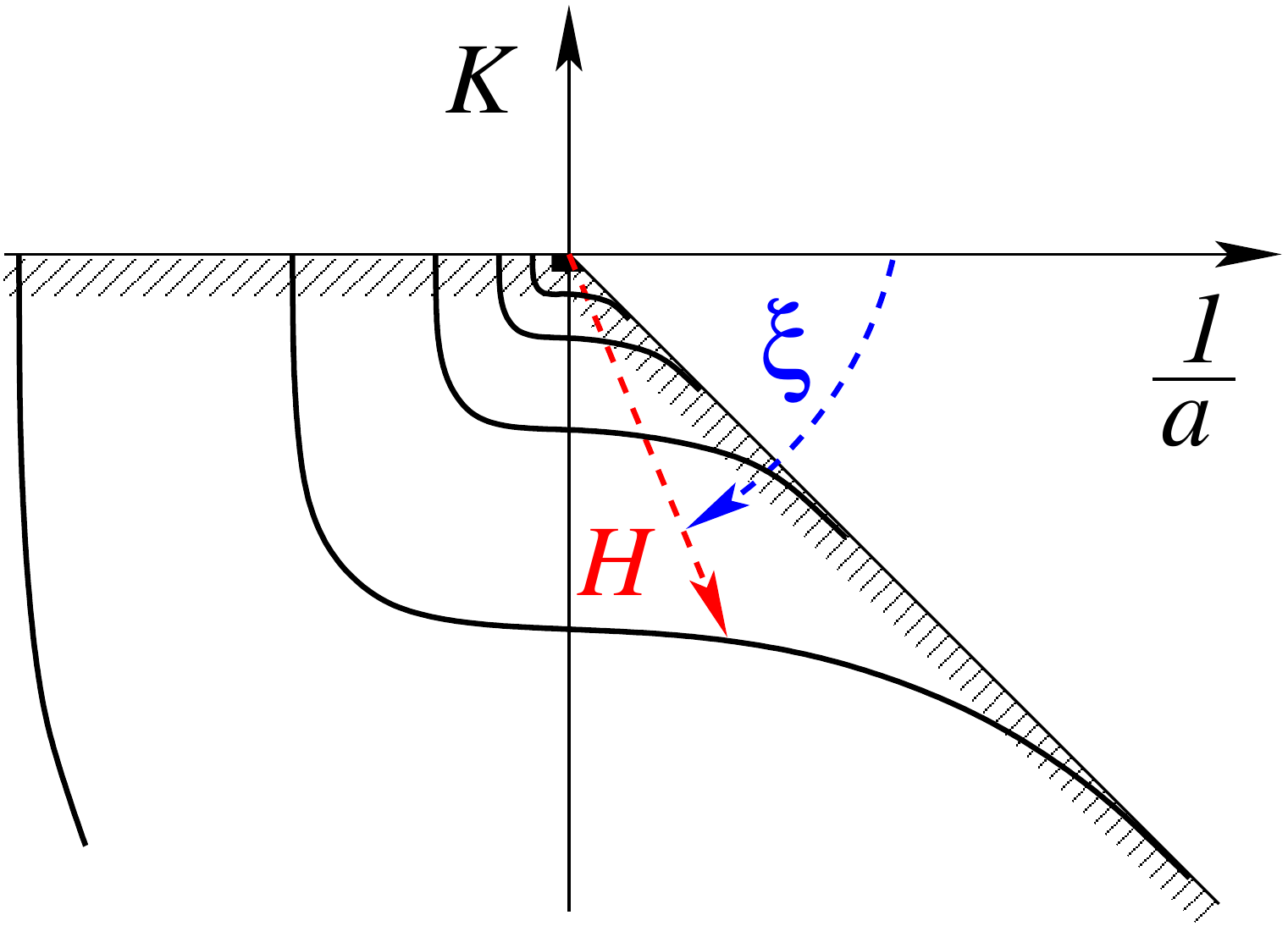}
  \caption{Illustration of the Efimov spectrum in the  $1/a-K$ plane.
The solid lines indicate the Efimov
states, while the hashed areas give the scattering thresholds.
The new variables $H$ and $\xi$ are illustrated by the dashed lines.
}\label{fig:efiplot}
\end{figure}

Efimov's equation can be derived by constructing a solution to
Eq.~(\ref{radial}) in the region $l \ll R \ll |a|$. We
define the radial variable $H$ and the angular variable $\xi$ via
\beq
H^2=mE_3^{(n)}/\hbar^2 +1/a^2\,\qquad\mbox{ and }\qquad
\tan\xi  = -a\sqrt {mE_3^{(n)}}/\hbar\,.
\label{Hxi-def}
\eeq
These variables are illustrated in Fig.~\ref{fig:efiplot} together
with the general form of the Efimov spectrum in the $1/a-K$ plane.
Since we are interested in low energies $mE_3^{(n)}/\hbar^2 \sim 1/a$,
the energy eigenvalue in (\ref{radial}) can be neglected. The
most general solution is a superposition of
outgoing and incoming hyperspherical waves~\cite{Efi71},
\beq
f (R) = \sqrt{H R} \left[A e^{is_0 \ln (H R)} + B e^{-is_0 \ln (H R)}
\right],
\label{f-general}
\eeq
where
the dimensionless coefficients $A$ and $B$ can depend on $\xi$.
At shorter distances $R \sim l$ and longer distances $R\sim |a|$,
the wavefunction is more complicated, but a solution
in this region is not required because of unitarity \cite{Efi71}.

We first consider the short-distance region. If there are
no deep two-body bound states with binding energies $E_2 \gg \hbar^2/(m a^2)$,
the two-body potential supports no bound states at all if $a<0$ and only the
shallow dimer with binding energy $E_2=\hbar^2/(ma^2)$ if $a>0$.
The probability in the incoming
wave must then be fully reflected at short distances and
we can set $B=Ae^{i\theta}$. The phase $\theta$ can be specified by giving the
logarithmic derivative of the radial wave function,
$R_0 f' (R_0)/ f (R_0)$, at any
point $l \ll R_0 \ll |a|$. The resulting expression for $\theta$
has a simple dependence on $H$:
\bqa
\theta/2 &=&s_0 \ln (H/ c\kappa_*)\,.
\label{theta-star}
\eqa
The quantity $c\kappa_*$ is a complicated function of
$R_0$ and $R_0 f'(R_0)/f(R_0)$. It differs by an unknown
constant $c$ from the three-body parameter $\kappa_*$
defined by Eq.~(\ref{B3-Efimov}) below.

We next consider large distances $R\sim |a|$. In general,
an outgoing hyperspherical wave incident on the $R\sim |a|$ region
can either be reflected or else transmitted to $R\to \infty$
as a scattering state.
For bound states in the region $-\pi < \xi < -\pi/4$, the probability must
be totally reflected such that $B=Ae^{i\Delta(\xi)}$, where the phase
$\Delta$ depends on the angle $\xi$. Compatibility with the
constraint from short distances requires
$\theta = \Delta (\xi)  \; {\rm mod} \; 2\pi$.
Using Eq.~(\ref{theta-star}) for $\theta$ and inserting the
expression for $H$, we obtain Efimov's equation \cite{Efi71}
\beq
E^{(n)}_3  + {\hbar^2 \over ma^2} = \frac{\hbar^2 \kappa_*^2}{m}\,
e^{-2 \pi n/ s_0} \exp \left[\Delta \left( \xi \right)/s_0
\right]\,,
\label{B3-Efimov}
\eeq
where the
constant $c$ was absorbed into $\Delta(\xi)$ such that
$\Delta(-\pi/2)\equiv 0$. This convention is used throughout the paper.
Note that we measure  $E_3$ from the three-boson threshold. $\kappa_*$ is thus
the binding momentum of the state with label $n=0$ in the unitary limit.
Once the universal
function $\Delta(\xi)$ is known, the Efimov energies $E^{(n)}_3$
can be calculated by solving Eq.~(\ref{B3-Efimov}) for different
integers $n$.
This equation has an exact discrete scaling symmetry: if there is an
Efimov state with binding energy $E^{(n)}_3$ for the parameters $a$
and $\kappa_*$, then there is also an Efimov state with binding energy
$\lambda^2 E^{(n)}_3$ for the parameters $\lambda^{-1} a$ and $\kappa_*$
if $\lambda=e^{k\pi/s_0}$
with $k$ an integer.

Equivalently, Eq.~(\ref{B3-Efimov}) can be written in parametric form
as~\cite{Kievsky:2012ss}
\beq
\frac{E_3^{(n)}\,ma^2}{\hbar^2}=\tan^2\xi\,,\qquad \qquad
\kappa_* a = e^{n\pi/s_0} \frac{ e^{-\Delta \left( \xi \right)/(2s_0)}}{\cos\xi}\,.
\label{B3-Efimov-parametric}
\eeq
This form can be generalized straightforwardly
to include range corrections but the universal function
$\Delta(\xi)$ remains the same.
$\Delta(\xi)$ can  be calculated by solving the three-body problem
for the Efimov binding energies in various potentials whose scattering
lengths are so large that effective range corrections are negligible
or by solving the Skorniakov-Ter-Martirosian
integral equations for the zero-range case \cite{STM57,Braaten:2004rn}.

\section{Universal Function $\Delta(\xi)$}

An explicit
parametrization of the universal function \(\dxi\) was first extracted from
solutions of the zero-range STM equations derived from effective field theory
in Refs.~\cite{Braaten:2002sr,Braaten:2004rn}.\footnote{Note that the
  parametrizations in Refs.~\cite{Braaten:2002sr,Braaten:2004rn} are
  equivalent. However, in \cite{Braaten:2004rn}, the universal function
  \(\dxi\) was shifted by the overall constant 8.22 such that
  $\Delta(-\pi/2)\equiv 0$.}
We start by revisiting the parametrization
from \cite{Braaten:2004rn}, which has the general form
\begin{equation}
  \Delta(\xi)=P_f^{(n)}\left( \xi \right)
  \equiv \sum_{k=0}^{n} c_k \cdot \left( f\K{\xi} \right)^k
  \label{eq:polynom_adv}\,.
\end{equation}
The parametrization was given separately for the three intervals
$\pida < \xi \leq \piv$, $\pifa < \xi \leq \pida$,
and $-\pi \leq \xi \leq \pifa$. It reads
\begin{align}
  \Delta(\xi)&=
  \begin{cases}
    3.10 \, f_1(\xi)^2 - 9.63 \, f_1(\xi) +6.04\,, & \pida < \xi \leq \piv\,, \\
    1.17 \, f_2(\xi)^3 +1.97 \, f_2(\xi)^2 +2.12 \, f_2(\xi)\,, & \pifa < \xi \leq \pida\,, \\
    0.25 \, f_3(\xi)^2 +0.28 \, f_3(\xi) -0.89 \,,  & -\pi \leq \xi \leq \pifa\,,
  \end{cases}
  \label{eq:oldpara}
\end{align}
where
\begin{align}
  f_1\left(\xi\right) &= \sqrt{-\frac{\pi}{4}-\xi},
  \label{eq:fit_expansion_variable}\\
	f_2\left(\xi\right) &= \frac{\pi}{2}+\xi, \\
	f_3\left(\xi\right) &= \left( \pi + \xi \right)^2 \cdot \exp\left(-\left( \pi + \xi \right)^{-2}\right)\,.
\end{align}

\begin{figure}[H]
  \centerline{
    \includegraphics[width=0.5\textwidth]{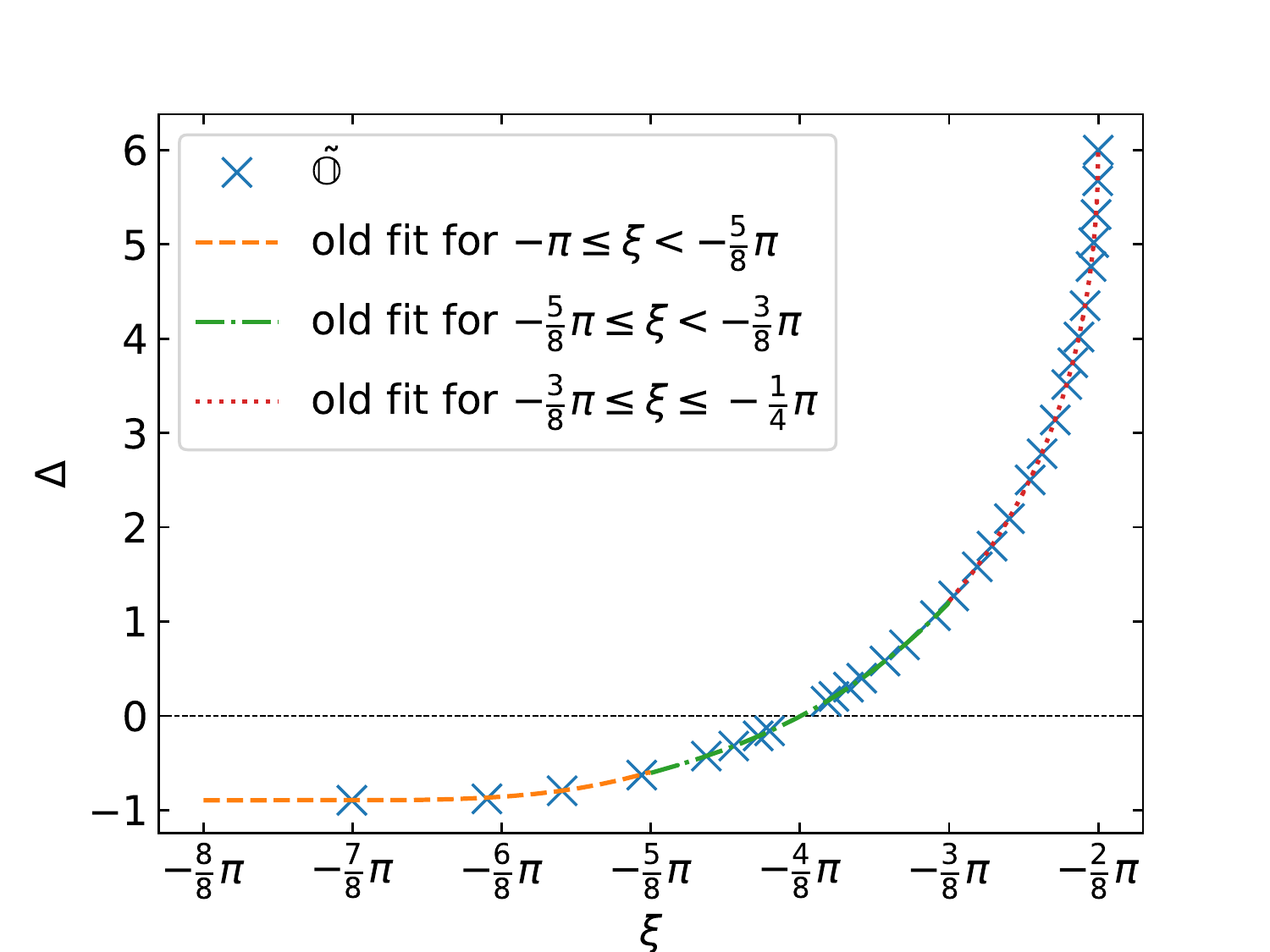}
    \includegraphics[width=0.5\textwidth]{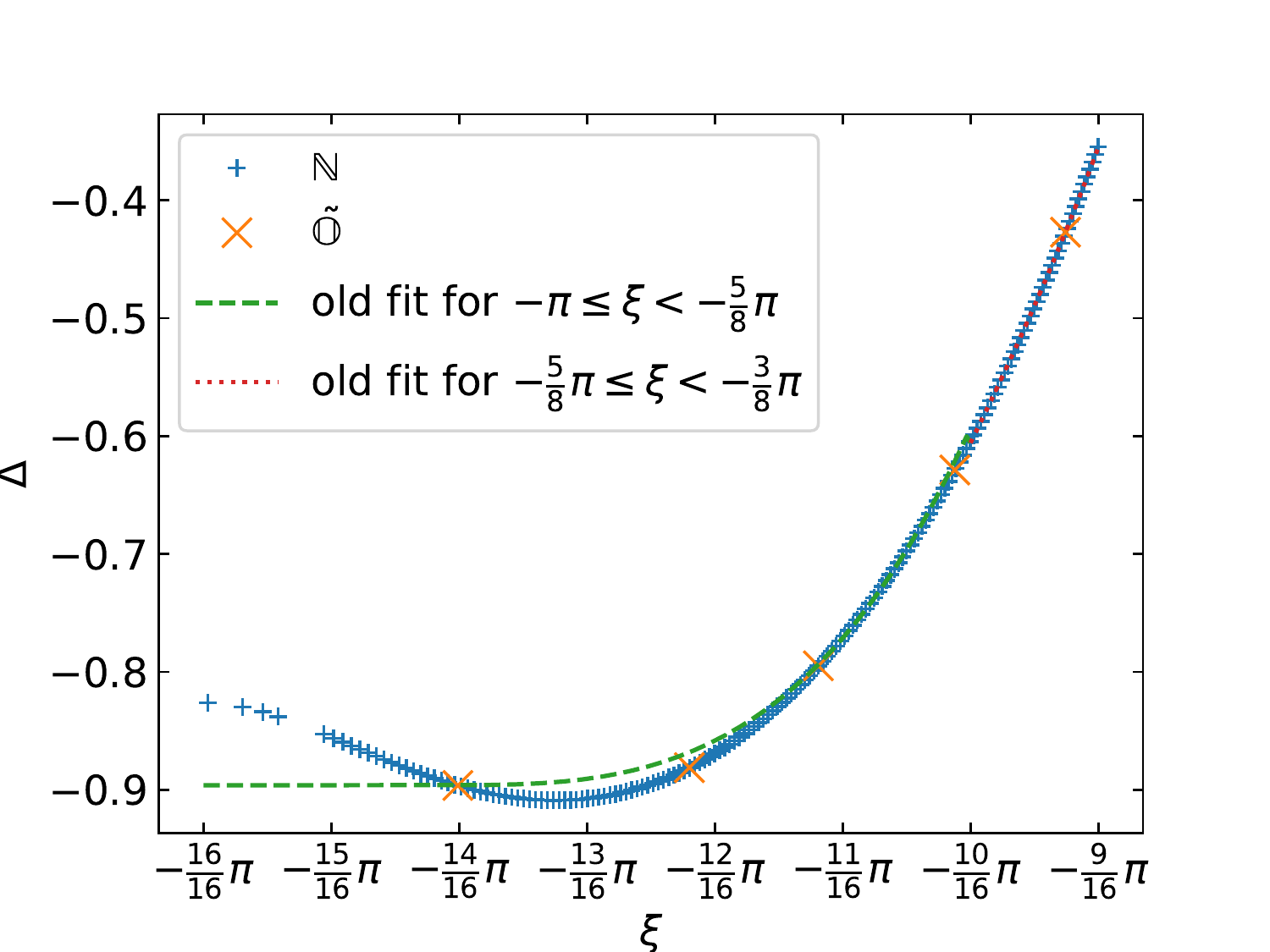}}
  \caption{Left panel: parametrization (\ref{eq:oldpara}) with old data set
    from \cite{Braaten:2002sr},
    which was shifted approximately by 8.214 in order to use the same convention
    as \cite{Braaten:2004rn}.
    Right panel: parametrization (\ref{eq:oldpara})
    with shifted old data set \sods~and new data set \nds~in the region $\xi \simge -\pi$.}
  \label{fig:oldpara}
\end{figure}

In the left panel of Fig.~\ref{fig:oldpara}, we show the parametrization of
Eq.~(\ref{eq:oldpara}) together with an shifted version of the original data set \ods,
which consists
of 31 data pairs $(\xi, \Delta(\xi))$ in total. The data are well captured by this
parametrization.
We have obtained a new set of binding energies $E_3$ as a function of the
inverse scattering length $1/a$ by numerically
solving the STM integral equation with a cutoff in momentum space
\cite{STM57}. In particular, we take Eq.~(340) of Ref.~\cite{Braaten:2004rn}
for $H(\Lambda)=0$,
\begin{equation}
  {\cal B}(p)  = \frac{4}{\pi}\int_0^\Lambda dq\,q^2\left[ \frac{1}{2pq}
  \ln\frac{p^2+pq+q^2 - E_3 -i\epsilon}{p^2-pq+q^2 - E_3 -i\epsilon}    \right]
  \left[-\frac{1}{a} + \sqrt{\frac{3q^2}{4} - E_3 - i\epsilon} \right]^{-1}
  {\cal B}(q) \,,
  \label{eq:integralEquation}
\end{equation}
where ${\cal B}(p)$ gives the residue at the pole of particle-dimer scattering
amplitude at the trimer energy $E_3$, and $\Lambda$ is the cutoff.  In
principle there is an additional three-body term $H(\Lambda)$
with log-periodic behaviour imposed by discrete scale
invariance~\cite{BHK99}. In the present calculations we just
set it to zero;  by doing so, the cutoff $\Lambda$ plays the role of
the three-body parameter \cite{Khar73,Hammer:2000nf}.
Thus $\Lambda$ is the scale that fixes the
energies, $E_3= \hbar^2\kappa^2/m\Lambda^2$, and the lengths, $a = \tilde
a/\Lambda$. To get rid of the finite cutoff effects, we need to consider
high excited states; in the present calculations we have calculated the second
excited state.

In order to numerically solve Eq.~(\ref{eq:integralEquation}), we have chosen a
grid for the values of the dimensionless momenta,
$q/\Lambda$ and $p/\Lambda$, turning Eq.~(\ref{eq:integralEquation}) into
an eigenvalue problem
\begin{equation}
  {\cal B} = {\cal I}(\kappa^2) {\cal B}
  \label{eq:integralMatrix}
\end{equation}
for the matrix ${\cal I}(\kappa^2)$ which is a function of the three-body
energy $E_3= \hbar^2\kappa^2/m\Lambda^2$; the value sought for the trimer
energy is that for which  the eigenvalue is equal to one.

In Table~\ref{tab:specialValues}, we report our values of the
binding momentum $\kappa_*$ at the unitary point,
our values of scattering lengths at the three-particle, $\kappa_*a_-$, and
particle-dimer, $\kappa_*a_*$, thresholds in units of $\kappa_*$ for the
first few states together with the ratio of the momenta.
To assess the accuracy of our calculation,
we compare our values with the theoretical prediction of
Refs.~\cite{Gogolin:2008xx, Braaten:2004rn}.

\begin{table}[h]
\begin{tabular}{c|c|c|c|c}
Level & $\kappa_*^{(n)}$ & $\kappa_*^{(n)}/\kappa_*^{(n+1)}$ &  $\kappa_*^{(n)} a_-^{(n)} $ & $\kappa_*^{(n)} a_*^{(n)} $ \\
\hline
\hline
0 & $1.7793756\cdot 10^{-1}$ & 22.9310 & -1.4485 & 0.055336 \\
1 & $7.7597079\cdot 10^{-3}$ & 22.6948 & -1.5044 & 0.073710 \\
2 & $3.4191512\cdot 10^{-4}$ & 22.6940 & -1.5069 & 0.070907 \\
3 & $1.5066314\cdot 10^{-5}$ &                    &                   \\
$\vdots$ & $\vdots$ &$\vdots$ &$\vdots$ &$\vdots$ \\
$\infty$ & 0                     & 22.69438 &
-1.50763~\cite{Gogolin:2008xx}& 0.07076~\cite{Braaten:2004rn}
\end{tabular}
\caption{Some special values obtained by numerically solving Eq.~(\ref{eq:integralEquation}).}
\label{tab:specialValues}
\end{table}

The new data set contains 5029 pairs $(1/a, E_3)$
which are given at 3-4 digit accuracy. In order to determine the
universal function $\Delta(\xi)$, the data were converted to pairs
$(\xi, \Delta(\xi))$ using Eq.~(\ref{B3-Efimov-parametric}).
Thus the new data set \nds~is both more accurate and more comprehensive
than the data set used in Refs.~\cite{Braaten:2002sr,Braaten:2004rn}.
In the right panel of  Fig.~\ref{fig:oldpara}, we show the parametrization
of Eq.~(\ref{eq:oldpara}) together with this new data set in the region
$\xi \simge -\pi$.
While the two data sets agree in this region, there is
clearly some structure that was not captured in Eq.~(\ref{eq:oldpara})
due to the limited number of data points.

Furthermore, the parametrization~(\ref{eq:oldpara}) is
only approximately continuous at the endpoints of the three
intervals. The resulting discontinuity of $\Delta(\xi)$
at $-3\pi/8$ ($-5\pi/8$) is 0.015 (0.013),
while the discontinuity of the first derivative is
0.38 (0.10), respectively.
The maximum absolute deviation with respect to the used data set is \(1.30 \cdot 10^{-2}\).
Finally, the polynomial $f_3(\xi)$ has an
essential singularity at $\xi=-\pi$ which is not mandated
by the underlying physics.
An updated version of this three-piece parametrization
without the essential singularity that improves
the continuity was recently given in Ref.~\cite{Naidon:2016dpf}.
Its discontinuity
at $-3\pi/8$ ($-5\pi/8$) is 0.0047 (0.0048), 
and the discontinuity of the first derivative is
0.26 (0.25), respectively. 
The aim of the present paper is to provide a new, accurate, and covenient
parametrization of Efimov's universal function that is continuous
and continuously differentiable over the whole interval.

\section{New Parametrization for $\Delta(\xi)$}

In order to obtain an optimal parametrization based on the new data \nds,
we have carried out analytical and numerical studies of
expansions of $\Delta(\xi)$ around the points
\(-\pi\), \(-\pi/2\), and \(-\pi/4\).
We chose these points because of their physical meaning.
In particular, we have
investigated different fits of parametrizations
consisting of one, two, and three
pieces with the above expansion points. For multiple piece fits,
different algorithms to insure the continuity of the function and its
derivative at the endpoints were implemented.\footnote{The difference of the
  two algorithms is that one constructs a basis of global
  continuously differentiable functions and does a global fit, while the other
  does the fits for each part consecutively ensuring the
  continuously differentiable connection each time.
  In this case the result generally depends on the order
  in which the fits are carried out.}

We found that a single piece fit with an expansion around
\(-\pi/4\)  is well suited to describe the data with a
reasonable number of terms.\footnote{Note that multiple piece fits
  are in principle more efficient due to the smaller intervals. However,
  once exact continuity of the function and its derivative is enforced
  at the endpoints, this
  advantage disappears.}
Our final parametrization is based on a single piece fit based on
Eq.~(\ref{eq:polynom_adv})\,,
where the function \(f\), which defines the expansion variable, is
given by $f_1$ from Eq.~(\ref{eq:fit_expansion_variable}).
This corresponds to an expansion around $\xi=-\pi/4$.\footnote{As an alternative
  to parametrizing \(\dxi\), we have also
  considered direct parametrizations of \(\log{E_3}\).
  However, this approach did not lead to any improvements.}

Up to this point our analysis was based on linear least-square fits, which minimize the
sum of the squared deviations \(d_i^2\). However,
a minimum \(\chi^2 = \sum_i d_i^2\) does in general not imply a minimum of
the maximum of the absolute deviations
$d_{\rm max}= {\rm max}\K{\left\{|d_i|\right\}}$.
In fact, a minimization of \(\chi^m = \sum_i d_i^m\) with \(m>2\) should
yield smaller \maxDeviation.
Doing such minimizations with generic minimization algorithms appeared to be
challenging because of strong varying quality of results.
The solution to this problem is a fitting method
developed by
Lawson~\cite{Lawson:1961,Rice:1968}.
This algorithm minimizes \maxDeviation~by doing a series
of standard least-square fits.

In addition to the 5029 new data points discussed above our fit also
includes the known values of $\Delta(\xi)$
at the atom-dimer and three-particle thresholds, which form the data set \tds.
The value at the atom-dimer threshold,
\beq
\Delta(-\pi/4) \approx 6.02730678199\,,
\eeq
was taken
from Ref.~\cite{Braaten:2004rn}. The value at the three-particle threshold,
\beq
\Delta(-\pi) \approx -0.82619948\,,
\eeq
was calculated from the expression
\begin{equation}
  \label{eq:3pt}
	\Delta(-\pi) = - 2 s_0 \ln |a_*^\prime \kappa_*|
\end{equation}
using $s_0 \approx 1.0062378$ (calculated according to \cite{Braaten:2004rn}) 
and $a_*^\prime \kappa_* \approx -1.5076300$ 
which was calculated using \(a_*^\prime \kappa_* = -2 \exp \K{\pi\gamma/ s_0} \)
and \(\gamma \approx -0.090518155\)
from Ref.~\cite{Gogolin:2008xx}. The expression (\ref{eq:3pt})
can be derived by evaluating Eq.~(\ref{B3-Efimov}) for $n=0$
at the three-particle
threshold, i.e., setting
\(\xi=-\pi\), \(a=a_*^\prime\), and \(E_3^{(0)}=0\).
The value at $\xi =-\pi/2$: $\Delta(-\pi/2)=0$ was already included
in the new data set \nds. Thus our fit, which is based on \cds,
includes 5031 data points in total.
Unless otherwise noted, deviations are given with respect to this data set.
We have also performed an analysis concerning the consistency of \tds~with \nds,
further information can be found in Appendix \ref{sec:fitting_methodology}.

We have performed fits from \(6\) up to \(11\) coefficients.
The procedure was the following: (i) we performed a Lawson fit,
(ii) we rounded the obtained coefficients to a specified number of digits,
(iii) we optimized the rounding to minimize the deviation from the
data set.
After each step the deviation values were computed in order to evaluate the fit.
Details on the optimized rounding procedure can be found in
Appendix \ref{sec:opt-rounding}.

Our new parametrization has the form
\beq
\Delta(\xi)= \sum_{k=0}^{7} c_k \: \left(-\frac{\pi}{4}-\xi \right)^{k/2}
\label{eq:new_para}\,,
\eeq
where the coefficients $c_k$
are given in Table~\ref{tab:final_fit_coefficients}.
\begin{table}[htb]
  \begin{tabular}{c||cccccccc}
\(k\)\: & 0 & 1 & 2 & 3 & 4 & 5 & 6 & 7\\\hline
\(c_k\)\: & 6.027 & -9.75 & 5.56 & -12.75 & 27.77 & -29.29 & 15.06 & -3.01\\
\end{tabular}
   \caption{Coefficients $c_k$ of the new parametrization (\ref{eq:new_para})
    of $\Delta(\xi)$.}
  \label{tab:final_fit_coefficients}
\end{table}
The values of our new parametrization at the thresholds and at unitarity as well
as the value of the derivative at unitarity are given in
Table~\ref{tab:final_fit_values}.
\begin{table}[htb]
  \begin{tabular}{cccc}
 {\leftValue} & {\middleValue} & {\rightValue} & {\derivativeValue} \\
\hline
-0.8251 & 0.0003779 & 6.027  & 2.126 \\
  \end{tabular}
  \caption{Special values of the new parametrization.}
  \label{tab:final_fit_values}
\end{table}

\begin{figure}[htb]
  \centerline{
  \includegraphics[width=0.5\textwidth]{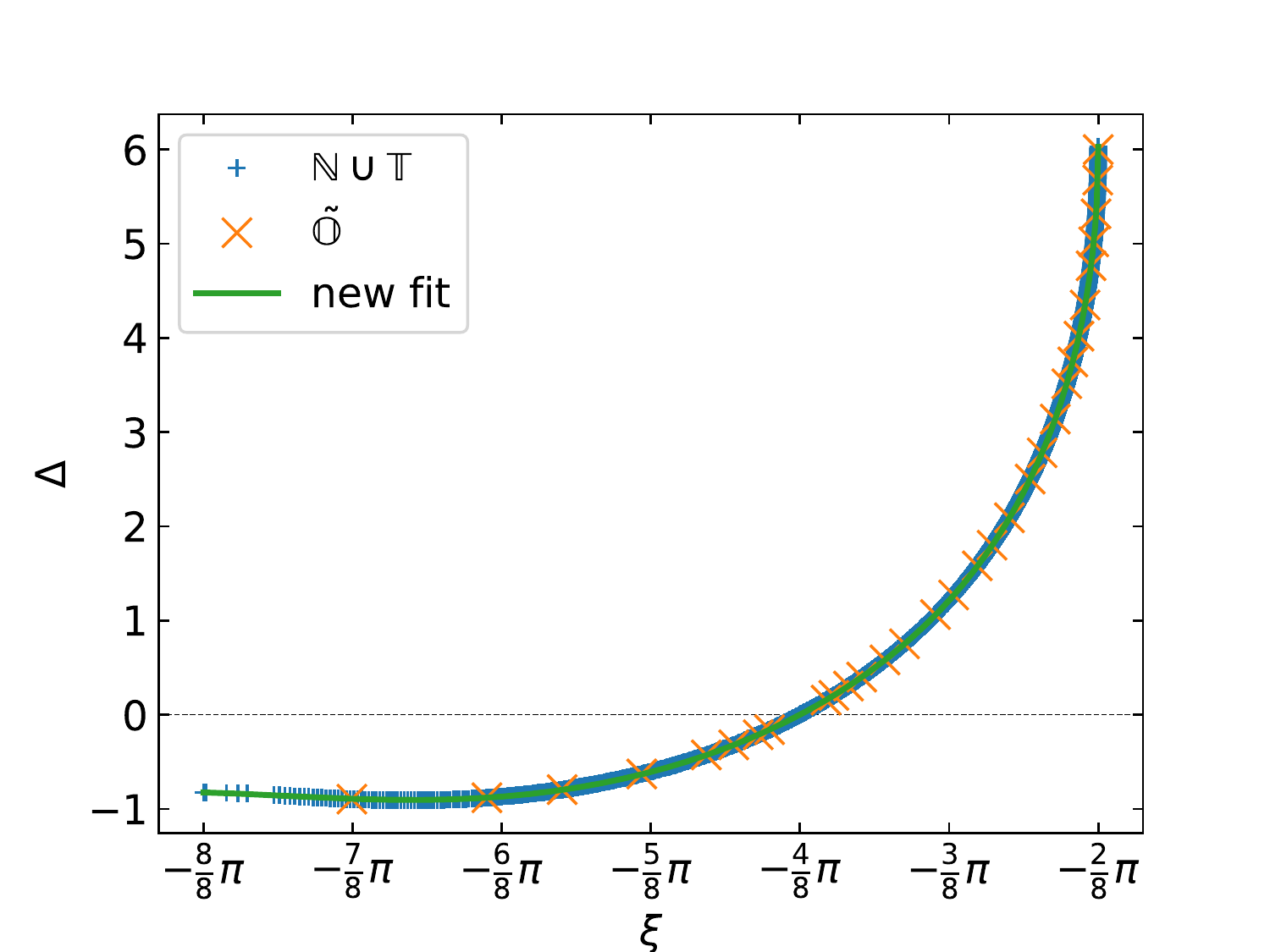}
  \includegraphics[width=0.5\textwidth]{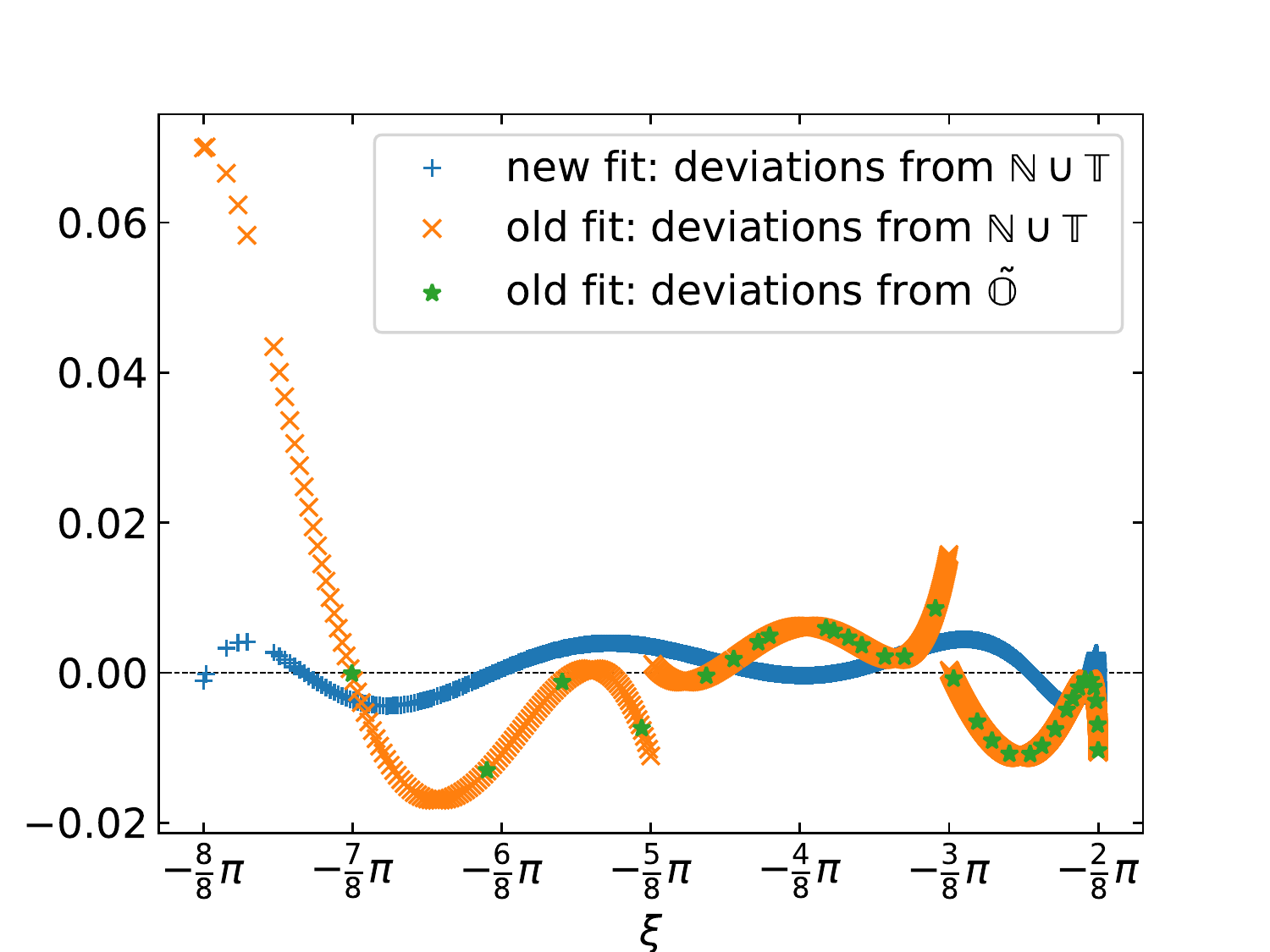}
  }
  \caption{Left panel: comparison of the new parametrization
    (\ref{eq:new_para}) with the
    new and old data sets. Right panel: deviations of the new
    parametrization (\ref{eq:new_para}) and the old parametrization
    from the new data set.}
  \label{fig:final_fit_all_data}
\end{figure}

In the left panel of Fig.~\ref{fig:final_fit_all_data}, we show
the new parametrization (\ref{eq:new_para}) in comparison with the new and
the old data sets. Both data sets are captured well by the new parametrization.
In the right panel of Fig.~\ref{fig:final_fit_all_data}, we show
the deviations of the two parametrizations from the new data set.
Clearly, the new parametrization has a maximum deviation that is about
one order of magnitude smaller and is smooth everywhere
by construction.
A more detailed comparison of the fits can be found in
Appendix~\ref{sec:fit_comparison}.
Better accuracy could be achieved by using more coefficients and/or
keeping more digits of the coefficients, but the parametrization
(\ref{eq:new_para}) is a good compromise between accuracy and
user friendliness.

\section{Range Corrections}

On the one hand,
Kievsky and Gattobigio have shown that the Efimov radial law, given in
Eq.~(\ref{B3-Efimov-parametric}),
can be generalized to include range corrections by making some minor
modifications~\cite{Kievsky:2012ss}:
\bqa
\frac{E_3^{(n)}}{E_2}&=&\tan^2\xi\,,\nonumber\\
\kappa^{(n)}_* a_B +\Gamma^{(n)}&=& \frac{ e^{-\Delta \left( \xi \right)/(2s_0)}}{\cos\xi}\,,
\label{B3-Efimov-parametric-range}
\eqa
where
\beq
\frac{1}{a_B} = \frac{\sqrt{mE_2}}{\hbar}\,
\eeq
is the binding momentum of the dimer and
$\kappa^{(n)}_*=\kappa_* e^{-n\pi/s_0}$. As in the zero-range case,
the sign of $a_B$ is chosen to be positive if the dimer is bound and
negative if the dimer is a virtual state.
The radial law now depends on two-parameters:
the finite-range parameter $\Gamma^{(n)}$ and
the three-body parameter $\kappa^{(n)}_*$,
both depending on the branch $n$.

Close to the unitary limit, Eq.~(\ref{B3-Efimov-parametric-range})
is well verified by a variety of potentials.
The specific value of the finite-range parameter depends very little on the
particular form of the potential~\cite{raquel2016} and can be estimated using
a two-parameter potential of Gaussian form.
Accordingly, using an S-wave Gaussian
interaction of range $r_0$, the values of the energy and range parameters
for the first three states are given in Table~\ref{tab:range_results}.
\begin{table}[htb]
  \begin{tabular}{c|ccc}
    $n$ & 0 & 1 & 2\\
    \hline
    $r_0 \kappa^{(n)}_*$ & 0.4874 & 0.02124 & 0.000915\\
    $\Gamma^{(n)}$ & 0.8472 & 0.06002 & 0.0035
  \end{tabular}
  \caption{The three-body parameter $\kappa^{(n)}_*$, in units of the Gaussian
    range $r_0$, and the finite-range parameter $\Gamma^{(n)}$
    for lowest three states.}
  \label{tab:range_results}
\end{table}

On the other hand, Ji et al.~\cite{Ji:2015hha}
have shown that the range corrections
can approximately be taken into account by introducing a running
three-body parameter:
\beq
\bar\kappa_* (\mu_0,a)\equiv(\mu_0/\kappa_*)^{-\gamma r_s/a}\kappa_*\,,
\label{eq:running_ks}
\eeq
where  $\gamma = 0.351..$, $\mu_0$ is a momentum scale,
and $r_s$ the effective range.
The running parameter modifies the zero-range parameter $\kappa_*$ at
each value of $a$ once range corrections are taken into account.
Large logarithms in the range corrections can be avoided by expressing
observables in terms of $\bar\kappa_*$.
For each level it corresponds to the three-body parameter
$\kappa^{(n)}_*$ defined above. Moreover
close to the unitary limit $a$ and $a_B$ $\rightarrow\infty$ and
$\bar\kappa_*\rightarrow \kappa_*$. The running three-body
parameter can be expanded around this point as
\beq
\kappa_*\approx \bar\kappa_*\left( 1+\frac{\gamma r_s}{a_B}\ln(\mu_0/\bar\kappa_*)\right)\,,
\label{eq:running_ks1}
\eeq
where the use of $a_B$ and $\bar\kappa_*$ in the small term is equivalent
at first order.

\begin{figure}[h]
    \includegraphics[width=0.7\textwidth]{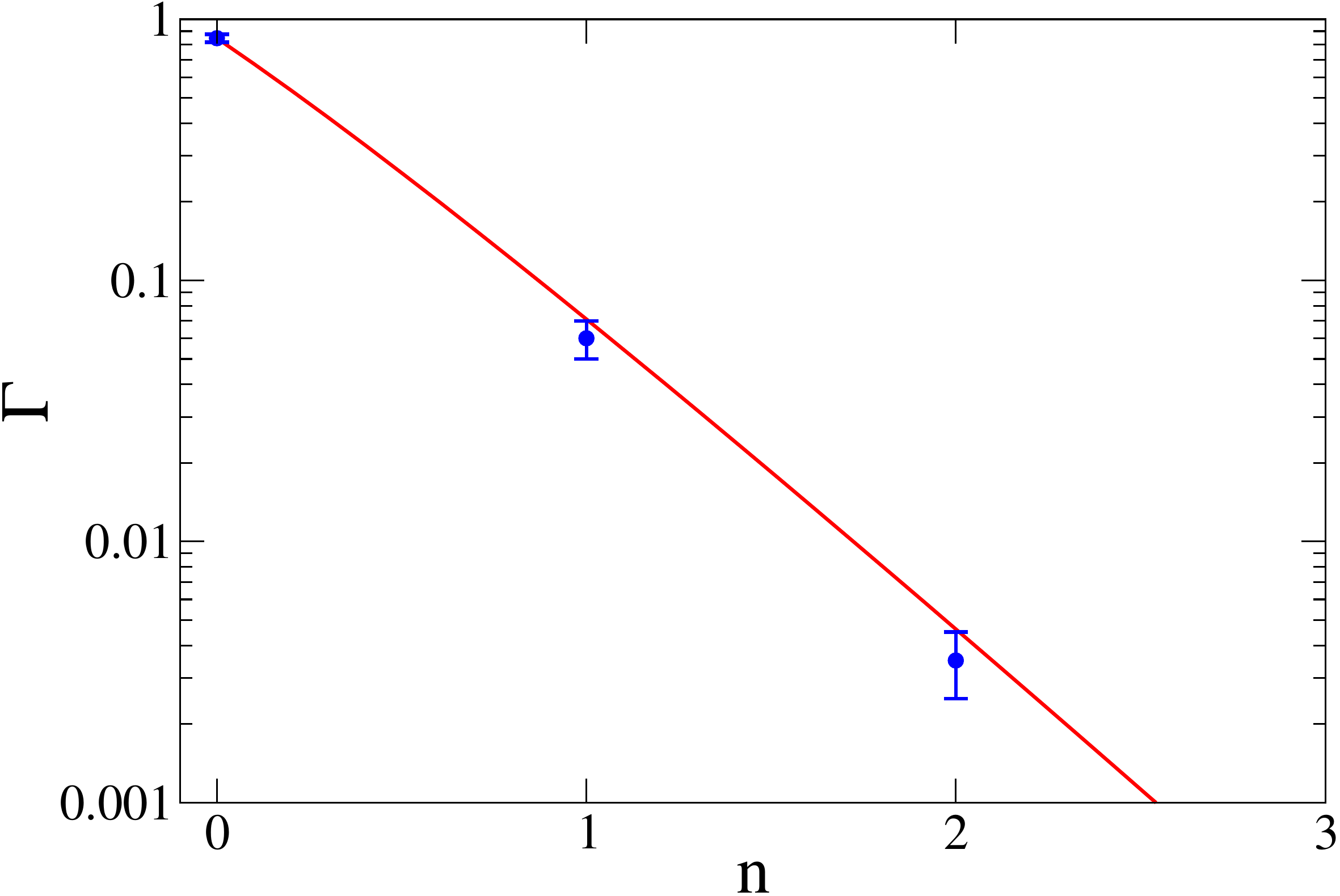}
  \caption{The finite-range parameter $\Gamma^{(n)}$ as a function of the level
    $n$. The solid line is given by Eq.~(\ref{eq:running_ks3}) with
    $\mu_0 r_0\approx 15.3$ whereas the
     circles are the values of Table~\ref{tab:range_results}. The error bars
     indicate the spread of different model potentials.}
  \label{fig:gamman}
\end{figure}

In  Eq.~(\ref{B3-Efimov-parametric-range}) the running parameter is given
in terms of
$\Gamma^{(n)}$.  The equation can be put in the form
\beq
e^{-n\pi/s_0}\kappa_* a_B = \kappa^{(n)}_* a_B + \Gamma^{(n)}\,,
\label{eq:running_ks0}
\eeq
where the binding momentum in the zero-range limit is
\beq
\kappa_* a_B = e^{n\pi/s_0} \frac{ e^{-\Delta \left( \xi \right)/(2s_0)}}{\cos\xi}\,,
\label{eq:running_ks2}
\eeq
Considering that Eq.~(\ref{eq:running_ks}) implicitely carries a factor
$e^{n\pi/s_0}$, we can make the following identification
\beq
\Gamma^{(n)}= \gamma r_s \frac{\bar\kappa_*}{e^{n\pi/s_0}}\ln{\K{\frac{\mu_0}{\bar\kappa_*}e^{n\pi/s_0}}}
= \gamma r_s \frac{\bar\kappa_*}{e^{n\pi/s_0}}\left( \ln{\K{\frac{\mu_0}{\bar\kappa_*}}}+ n\pi/s_0\right)\,.
\label{eq:running_ks3}
\eeq
This equation shows the functional dependence of the shift
$\Gamma^{(n)}$ in terms of
the level $n$ and the momentum scale $\mu_0$. For a Gaussian of range $r_0$
it can be put into the form
\beq
\Gamma^{(n)}= \gamma (r_s/r_0) \frac{\bar\kappa_* r_0}{e^{n\pi/s_0}}
\left( \ln{\K{\frac{\mu_0}{\bar\kappa_*}}}+ n\pi/s_0\right) \approx
\frac{0.2456}{e^{n\pi/s_0}} \left( 3.445 + n\pi/s_0\right)\,,
\label{eq:running_ks4}
\eeq
where we have used the ratio $r_s/r_0=1.43522$~\cite{raquel2016} and
the values of $r_0\kappa_*^{(n)}$ and $\Gamma^{(n)}$
for $n=0$ given in Table \ref{tab:range_results}
to determine $\mu_0 r_0\approx 15.36$. Using these values,
we can predict $\Gamma^{(1)}$ and $\Gamma^{(2)}$
from Eq.~(\ref{eq:running_ks4}). The corresponding evolution
of $\Gamma^{(n)}$ with $n$ is shown in Fig.~\ref{fig:gamman}.
Overall, we find good agreement with the values of
$\Gamma^{(1)}$ and $\Gamma^{(2)}$ in Table \ref{tab:range_results}
within the error bars given by the spread of different model potentials.
Finally, note that Eq.~(\ref{eq:running_ks4}) predicts a correction
to the naive expectation $\ln \Gamma^{(n)} = \mbox{const.} - n\pi/s_0$
which is confirmed by the finite range results from Gaussian potentials.

\section{Summary}

We have provided a new parametrization of
the universal function $\Delta(\xi)$ that appears in
Efimov's equation (\ref{B3-Efimov}, \ref{B3-Efimov-parametric})
for the binding energies $E_3^{(n)}$ of
Efimov states. This equation provides  a simple alternative
to the solution of the STM integral equations for the
calculation of universal Efimov binding energies.

In Ref.~\cite{Braaten:2002sr}, this equation was generalized
to include the effects of deep two-body bound states. The generalization
involves an additional inelasticity parameter $\eta$,
but the spectrum is determined by the same
universal function $\Delta(\xi)$ determined here.

A simple modification of Efimov's
universal equation, given in Eq.~(\ref{B3-Efimov-parametric-range}),
that can account for effective range corrections
was proposed in Ref.~\cite{Kievsky:2012ss}. It was shown to work
well for $^4$He atoms and other systems close to the
unitary limit.
We have quantitatively investigated the connection of this equation to
the running three-body parameter introduced by Ji et al.~\cite{Ji:2015hha}
in a rigorous perturbative treatment of effective range effects
and found good agreement. This result provides further evidence for
the validity of the finite range extension of Efimov's universal
equation proposed in Ref.~\cite{Kievsky:2012ss}.

\acknowledgments

We thank Eric Braaten, Wael Elkamhawy, and Fabian Hildenbrand
for discussions.
The work of HWH is supported by the Deutsche Forschungsgemeinschaft (DFG,
German Research Foundation) - Projektnummer 279384907 - SFB 1245 and
the Federal Ministry of Education and Research (BMBF) under contract
05P18RDFN1.

\appendix

\section{Details of the Fitting Methodology}
\label{sec:fitting_methodology}

Since plots of the new data set \nds~together with the values at the thresholds \tds~
questioned the consistency of the resulting data set \cds, we modified the first step of the fitting
procedure to address this question.
The single Lawson fit is replaced by a series of Lawson fits in the following way:
First a set of problematic data points from \nds~is defined (typically some points
around the thresholds from \tds). Then a Lawson fit without these problematic
points is carried out and the deviations of these data points from the resulting
function are calculated. Those data points whose deviation is smaller than
a given threshold
get included and a new fit is done.
This procedure is repeated until all points are included or no more points meet the
condition.

We used as threshold \(1.1 d_{\mathrm{max},0}\), where
\({d_{\mathrm{max},0}}\) is the maximum absolute deviation of the initial fit.
This algorithm had to be employed for all fits with a different number of coefficients.
The result of this procedure is that all fits
are based on the complete data set \cds~consisting of 5031 data points.

\section{Optimized Rounding Procedure}
\label{sec:opt-rounding}

The optimized rounding improves the quality of the parametrization
over standard rounding of the fit coefficients, especially when the
coefficients are rounded to a low number of decimal digits. Our procedure
was as follows: all coefficients \(c_i\), which are rounded to \(n_i\) digits,
were varied independently within a certain range with a step size of \(10^{-n_i}\)
in order to minimize \maxDeviation, which usually increases
by rounding.
This procedure was carried out for fits from \(6\) up to \(11\) coefficients,
which were rounded to two and three digits except for the zeroth coefficient.
It was rounded to higher number of digits (three or four), since
\(c_0=\Delta\K{-\pi/4}\) holds in case of our parametrization.
As a consequence we ended up with four different rounding schemes:
f3d2, f4d2, f3d3 and f4d3.
Here the notation f\(x\)d\(y\) is used with \(x\) as the number of decimal digits
of the zeroth coefficient \(c_0\) and \(y\) as number of decimal digits of the other
coefficients \(c_{i>0}\).
Thus in total \(2 \cdot 2 \cdot 6 = 24\) optimized rounding procedures had to
be carried out.
The interval in which the coefficients were varied in each optimized rounding
process was chosen so that in each process effectively about
\(10^{12}\) variations
were tested. This corresponds to 100 variations per coefficient in
a fit with 6 coefficients and 13 variations per coefficient in
a fit with 11 coefficients.

\begin{figure}[htb]
  \centerline{~~~
  \includegraphics[width=0.5\textwidth]{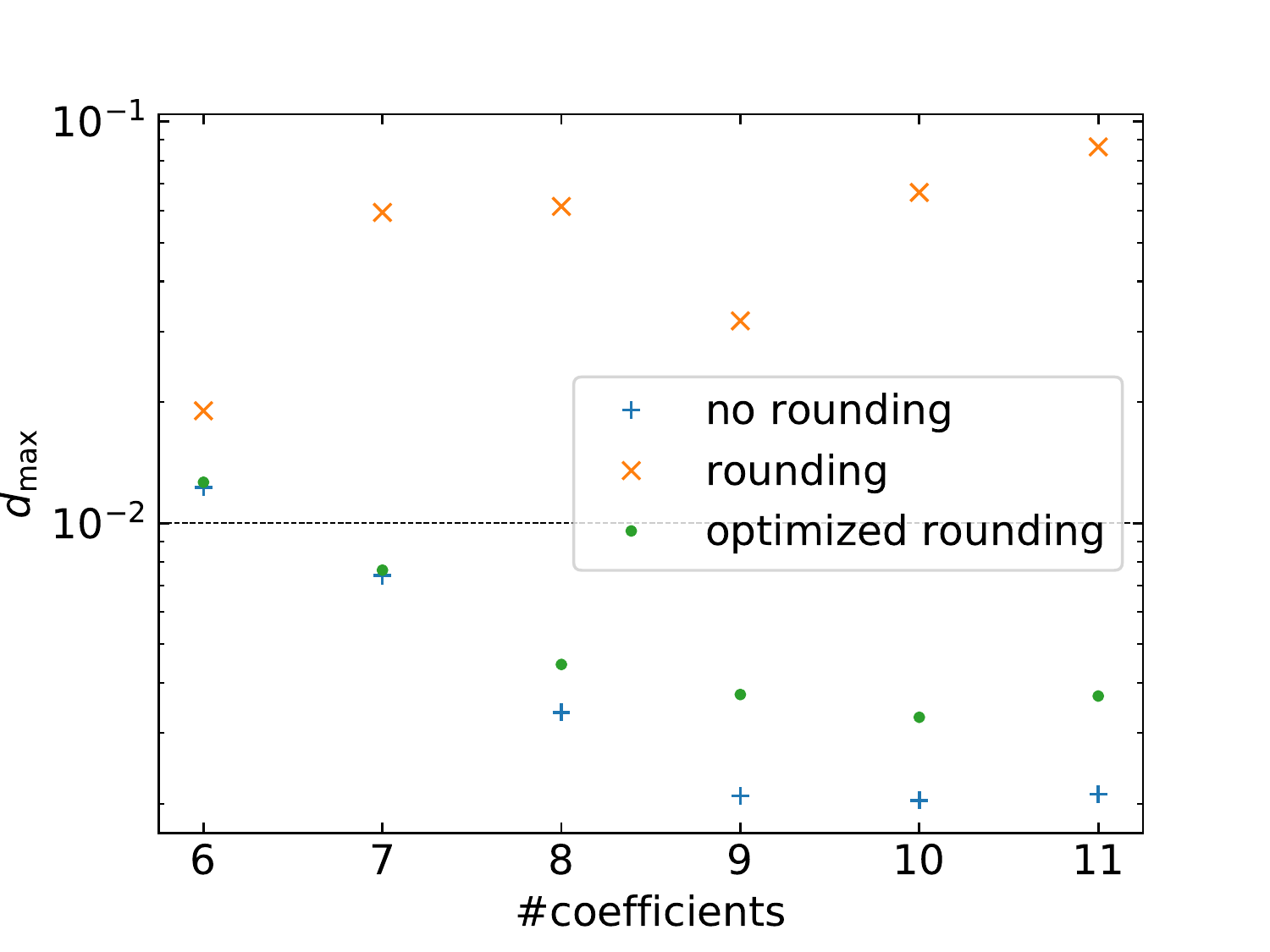}
  \includegraphics[width=0.5\textwidth]{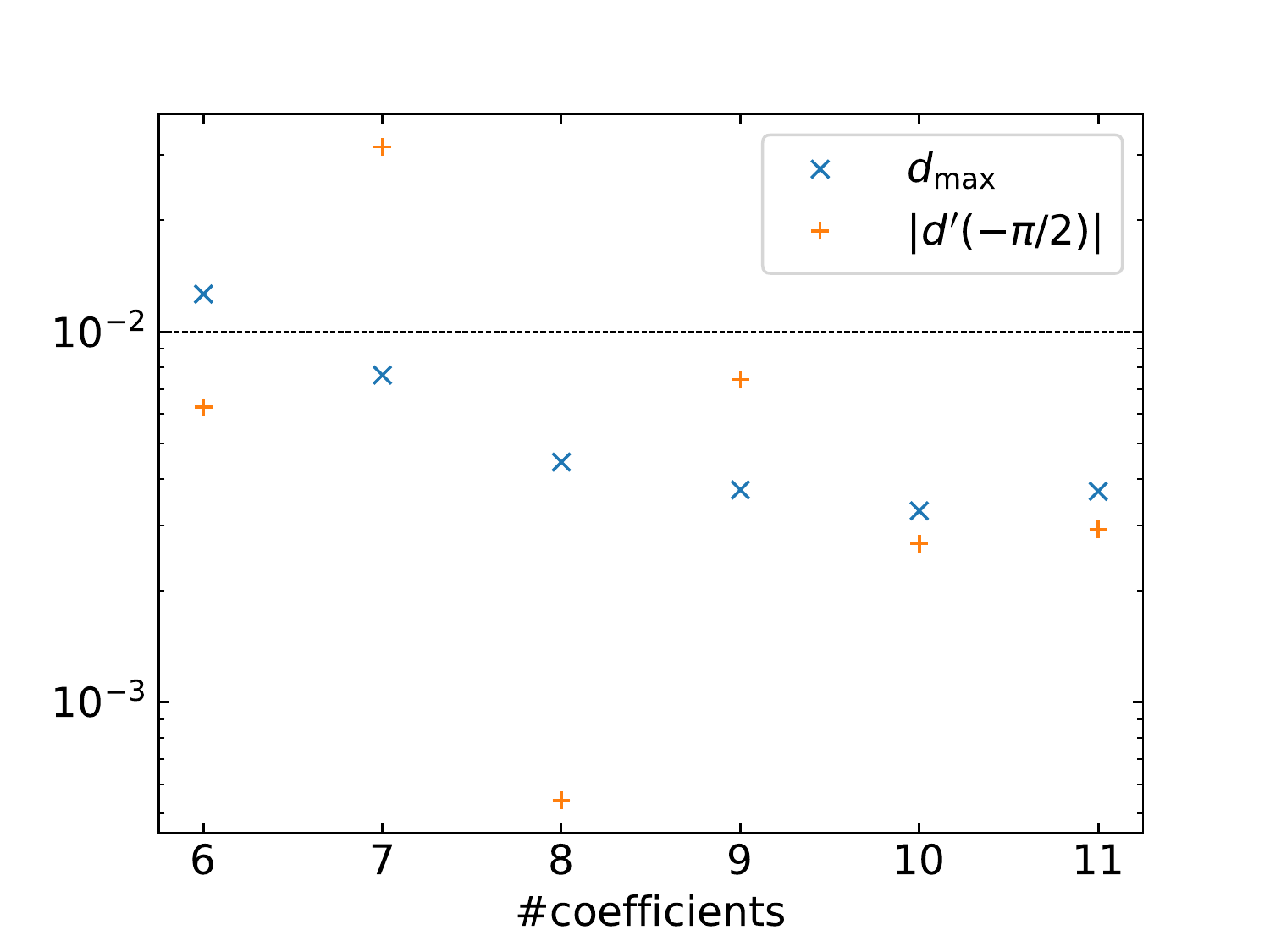}
  }
  \caption{Left panel: maximum absolute deviations \maxDeviation~of the f3d2 fits as
    function of the number of coefficients. Right Panel: \maxDeviation~and the
    deviation of the derivative at unitarity \derivativeDeviationShort~
    after optimized rounding in the f3d2 scheme.}
  \label{fig:or_max_deviations}
\end{figure}
In Fig.~\ref{fig:or_max_deviations}, we
show the deviations of the different fits.
These plots clearly show that the optimized rounding leads to a significant reduction
of the deviations.\footnote{As expected the deviations after optimized rounding
  are higher than the deviations
  before rounding, while the inverse case could occur due to the fact that the Lawson
  fit is an iterative method.}
We find that rounding the \(c_{i>0}\) to two decimal places
is enough to have an maximum absolute deviation smaller than \(10^{-2}\).
In this case at least seven coefficients are necessary.
In comparison with the fit with seven coefficients the fit with eight ones
has a lower \maxDeviation~and a deviation at \(-\pi/4\), which is smaller by
more than one order of magnitude.
Thus the fit with eight coefficients was chosen.
Another advantage of this fit is a much lower deviation of its derivative from
\(\Delta^\prime \K{-\pi/2}\), which is approximately given to \(2.125850069373\) in \cite{Castin:2011}.
These observations hold for the rounding schemes f3d2 and f4d2.
We selected f3d2, as the absolute deviation at \(-\pi/4\) of f4d2 is also greater
than \(10^{-4}\)
and this implies giving \(c_0\) with four decimal digits is not justifiable.
It should be a good compromise between accuracy and usability.
It is comparable to the complexity of the
old parametrization (\ref{eq:oldpara}).

\section{Comparison of Different Parametrizations}
\label{sec:fit_comparison}

In Table~\ref{tab:fit_comparison} the new parametrization is compared to the
parametrization from \cite{Braaten:2002sr} (respectively \cite{Braaten:2004rn})
and the one given in \cite{Naidon:2016dpf}.
\begin{table}[htb]
  \caption{Comparison of fits.
  One should note that we do not know which data set
  was used for the creation of the parametrization from \cite{Naidon:2016dpf}.
  \maxDeviation~is computed using \cds,
  \(d \K{\xi}\) denotes the deviation from \cds~at \(\xi\).
  \(d^\prime \K{\xi}\) denotes the deviation of the derivative at \(\xi\).}
  \label{tab:fit_comparison}
  \begin{tabular}{c|ccccc}
    Fit & \maxDeviation & \leftDeviationShort & \middleDeviationShort & \rightDeviationShort & \derivativeDeviationShort \\
    \hline
{Old fit} & 7.01 \(\cdot 10^{-2}\) & 6.99 \(\cdot 10^{-2}\) & 6.15 \(\cdot 10^{-3}\) & 6.55 \(\cdot 10^{-3}\) & 5.85 \(\cdot 10^{-3}\) \\
{Fit from \cite{Naidon:2016dpf}} & 8.64 \(\cdot 10^{-3}\) & 1.20 \(\cdot 10^{-3}\) & 0.00 \(\cdot 10^{0}\) & 3.07 \(\cdot 10^{-4}\) & 1.59 \(\cdot 10^{-2}\) \\
{New fit} & 4.44 \(\cdot 10^{-3}\) & 1.07 \(\cdot 10^{-3}\) & 3.78 \(\cdot 10^{-4}\) & 3.07 \(\cdot 10^{-4}\) & 5.42 \(\cdot 10^{-4}\) \\
  \end{tabular}
\end{table}
%

\end{document}